\documentstyle[12pt]{article}

\catcode`\@=11
\renewcommand{\thefootnote}{\fnsymbol{footnote}}
\newcommand{\beq}{\begin{equation}}
\newcommand{\eeq}{\end{equation}}
\newcommand{\ba}{\begin{array}}
\newcommand{\ea}{\end{array}}
\newcommand{\bea}{\begin{eqnarray}}
\newcommand{\eea}{\end{eqnarray}}
\newcommand{\bean}{\begin{eqnarray*}}
\newcommand{\eean}{\end{eqnarray*}}

\begin{document}

\begin{titlepage}

\hfill{DFPD 95/TH/05}

\hfill{hep-th/9502089}

\vspace{0.5cm}

\centerline{\LARGE{Algebraic-Geometrical Formulation of}}

\vspace{0.5cm}

\centerline{\LARGE{Two-Dimensional Quantum Gravity}\footnote[1]{Partly
supported by the European Community Research
Programme {\it Gauge Theories, applied supersymmetry and quantum
gravity}, contract SC1-CT92-0789.\\
e-mail: \\
bonelli@ipdgr4.pd.infn.it\\ marchetti@padova.infn.it \\ matone@padova.infn.it}}

\vspace{1.5cm}

\centerline{\Large{G.~Bonelli, P.A.~Marchetti and M.~Matone}}

\vspace{0.5cm}

\centerline{\large{Department of Physics ``G. Galilei'' - Istituto Nazionale di
Fisica Nucleare}}

\centerline{\large{University of Padova}}

\centerline{\large{Via Marzolo, 8 - 35131 Padova}}

\vspace{0.5cm}

\centerline{\large{ITALY}}

\vspace{2.5cm}

\centerline{ABSTRACT}

\vspace{0.5cm}

We find a volume form on moduli space of double-punctured Riemann surfaces
whose integral satisfies the Painlev\'e I recursion relations of the
genus expansion of the specific heat of 2D gravity. This allows us to express
the asymptotic expansion of the
specific heat as an integral on an infinite dimensional moduli space in the
spirit of Friedan-Shenker approach.
We outline a conjectural derivation of such recursion relations using the
Duistermaat-Heckman theorem.

\end{titlepage}

\setcounter{footnote}{0}

\renewcommand{\thefootnote}{\arabic{footnote}}

{\bf 1.}
Important results concerning 2D gravity appeared
in the last few years. Investigations of the measure
of Liouville
quantum gravity were considered in \cite{1} whereas nonperturbative
results have been obtained in the framework of matrix models
approach \cite{2} and in the topological approach in
\cite{3}\cite{Verlindes}\cite{Kontsevich}\cite{DiFrancescoItzyksonZuber}.
The relationships between the different
approaches to 2D gravity have been investigated (see \cite{rew} for
a review), in particular
it has been proved  \cite{matone1} that
the Liouville action enters
in the evaluation of intersection
numbers involved in topological gravity.
More recently it has been proved that
the specific heat of a model of pure gravity
can be expressed in terms of a series of integrals
on moduli spaces of punctured spheres
with the integrand involving the classical Liouville action
\cite{matone2}.
In \cite{BMM1} it has been shown that this model
corresponds to Liouville quantum gravity with a non trivial
$\Theta$-vacuum.

Interesting suggestions on the structure of Liouville path-integral
have been
considered in \cite{1}. However the explicit form of the measure
is still unknown.

Our aim in this paper is to start from the exact results
of matrix models and then to
write the contribution at genus $h$
to the asymptotic expansion of the
specific heat as an integral on
$\overline{\cal M}_{h,2}$, the
compactified
moduli spaces of double-punctured Riemann surfaces.
Let us shortly outline how this can be achieved.
It is well known that the specific heat can be identified with minus the
two-puncture correlator $\langle\left({\cal O}_0 \right)^2\rangle$
and this explains the appearence of $\overline{\cal M}_{h,2}$,
furthermore it
satisfies the Painlev\'e I (PI) equation.
One starts by considering the recursion relations of the asymptotic
expansion of PI. In order to reproduce them in terms of integrals
 on moduli spaces
one first notes that the Weil-Petersson two-form $\omega_{WP}$,
the natural K\"ahler form
on the moduli space of Riemann surfaces $\Sigma$,
has the remarkable property of satisfying the
{\it restriction phenomenon}. Namely, if we write \cite{Wolpert1}
$\omega_{WP}$ in terms of Fenchel-Nielsen coordinates $l_j,\tau_j$,
the restriction of
$$
\omega_{WP}=\sum_j dl_j\wedge d\tau_j,
$$
to the
submanifold $l_{k_1}=0,\ldots, l_{k_n}=0$, is the
sum of the Weil-Petersson two-forms on the moduli spaces
for the components of
the Riemann surface $\Sigma-\{{\rm nodes\; for\;}
k_1,\ldots,k_n\}$.
Second, one notes that the
structure of the recursion relations for the asymptotic expansion (see
(\ref{1})) appears to be
related to the structure of the boundary of the moduli space
$\overline{\cal M}_{h,2}$ involving Riemann surfaces with nodes
(see (\ref{miles})).
However, to obtain
such recursion relations  from the Weil-Petersson two-form
using Poincar\'e duality in the framework of algebraic geometry, one
meets the difficulty that the Poincar\'e dual
of $\omega_{WP}$
is not concentrated on the
boundary of moduli spaces. This problem can be solved by
introducing a volume form $\omega_{WP}^{3h-2}\wedge\omega_L$, where
the cohomology class $[\omega_L]$ of the two-form
$\omega_L$
 is the Poincar\'e dual of a divisor $D_L$ concentrated at the boundary
of $\overline{\cal M}_{h,2}$ and this permits the use of the restriction
phenomenon property of $\omega_{WP}$.
The choice of $D_L$ is quite
crucial and is determined by the structure of the theory.
This algebraic-geometrical
formulation derived directly from the matrix model strengths the
connection between topological and Liouville gravity.

Recovering the volume form on $\overline{\cal M}_{h,2}$ associated
to the PI, allows us to write the asymptotic series
as a unique integral of a suitable volume form on an infinite
dimensional space of double-punctured Riemann surfaces, in the spirit
of the Friedan-Shenker program for 2D quantum gravity.
This integral is the asymptotic counterpart of
an analogous integral on an infinite dimensional space of all
punctured Riemann spheres considered in \cite{BMM1} representing the
specific heat in the strong coupling region.

\vspace{0.5cm}

{\bf 2.} Let us shortly recall some basic facts about
the moduli space of stable curves
$\overline{\cal M}_{h}$, the Deligne-Knudsen-Mumford
compactification of moduli space, and the specific heat of 2D gravity.

 $\overline{\cal M}_{h}$  is a projective
variety and its boundary
$\partial \overline{\cal M}_{h}
=\overline{\cal M}_{h}\backslash {\cal M}_h$,
called the compactification
divisor, decomposes into a union of divisors
${D}_0,\ldots,{D}_{[h/2]}$ which are
 subvarieties of complex codimension one.
 A Riemann surface $\Sigma$ belongs to
${D}_k\cong \overline{\cal M}_{h-k,1}\times
\overline{\cal M}_{k,1}$, $k>0$ if it
has one node separating it into two components of genus $k$
and $h-k$. The locus in ${D}_0\cong \overline{\cal M}_{h-1,2}$
consists of surfaces that become, on removal of the node,
genus $h-1$ double punctured surfaces.
Surfaces with multiple nodes lie in the intersections
of the $D_k$'s.
The compactified moduli space $\overline{\cal M}_{h,n}$ of
Riemann surfaces with $n$-punctures $z_1,\ldots,z_n$
 is defined analogously to $\overline{\cal M}_h$. In particular
for $n=2$ we have
\beq
\partial \overline{\cal M}_{h+1,2}=
e_0^{(h+1)} \overline{\cal M}_{h,4}+\sum_{k=1}^h
e_k^{(h+1)}\overline{\cal M}_{h-k+1,2}\times\overline{\cal M}_{k,2}
+\sum_{k=0}^h
o_k^{(h+1)}\overline{\cal M}_{h-k+1,1}\times\overline{\cal M}_{k,3},
\label{miles}\eeq
in the sense of cycles on orbifolds, where the coefficients are
combinatorial factors.

The specific heat of pure gravity satisfies the PI,
${\cal Z}^2(t)-{\cal Z}''(t)/3=t$, where $t\equiv
\lambda_{1}^Rg_s^{-4/5}$, with $\lambda_{1}^R$ the renormalized
cosmological constant and $g_s$ the string coupling constant.
The asymptotic (genus) expansion of ${\cal Z}(t)$ has the form
\beq
{\cal Z}(t)\sim  \sum_{h=0}^\infty e^{2i\Theta(1-h)}
Z_h t^{-5h/2+1/2},\qquad t\to +\infty,
\label{0}\eeq
with the coefficients $Z_h$ satisfying the recursion relations
\begin{equation}
Z_0=1, \quad Z_{h+1}=
{25h^2-1\over 24}Z_h-{1\over 2}\sum_{k=1}^h Z_{h-k+1}Z_k,
\qquad \Theta=0,\pi/2,
\label{1}\end{equation}
where $\Theta=0$ corresponds to ordinary 2D gravity
\cite{2}, whereas $\Theta=\pi/2$ corresponds to the model with a non
trivial $\Theta$-vacuum
proposed in \cite{BMM1}.

In terms of the free energy $F$, the $n$-puncture correlator
is given by \cite{2}
\beq
\langle \left( {\cal O}_0\right)^n\rangle=
-{d^nF(t)\over dt^n},
\label{m1}\eeq
and ${\cal Z}(t)=F''(t)$.

For future purpose we write down the genus expansion
of the
4-puncture correlator
\beq
\langle \left( {\cal O}_0\right)^4\rangle=
{d^2\over dt^2}\langle \left( {\cal O}_0\right)^2\rangle\sim
-\sum_{h=0}^\infty e^{2i\Theta(1-h)}W_ht^{-5h/2 -3/2},\quad t\to
+\infty,\qquad
W_h={25h^2-1\over 4}Z_h.
\label{m2}\eeq

\vspace{0.5cm}

{\bf 3.} We now find a representation of $Z_h$ in the form
$\int_{\overline{\cal M}_{h,2}}\omega^{(h)}$,
with $\omega^{(h)}$ a $(6h-2)$-form on $\overline{\cal M}_{h,2}$.
We assume $\omega^{(h)}\equiv
\left(\omega_{WP}^{(h,2)}\right)^{3h-2}\wedge\omega_L^{(h)}/(3h-1)!$
with $\omega_L^{(h)}$ a two-form whose explicit expression
will be given later and $\omega_{WP}^{(h,n)}$ the Weil-Petersson
two-form on $\overline{\cal M}_{h,n}$ divided by $\pi^2$. We set
\beq
Z_h={1\over (3h-1)!}\int_{\overline{\cal M}_{h,2}}
\left(\omega_{WP}^{(h,2)}\right)^{3h-2}\wedge\omega_L^{(h)},\qquad h>0,
\label{2}\eeq
with initial condition $Z_0=1$.

We now define a divisor $D_L^{(h)}$, which we call ``Liouville
divisor'', as
the $(6h-4)$-cycle
\beq
D_L^{(h)}=
c_0^{(h)}\overline{\cal M}_{h-1,4}+
\sum_{k=1}^{h-1}c_k^{(h)}
\overline{\cal M}_{h-k,2}\times \overline{\cal M}_{k,2},
\label{4}\eeq
where the coefficients $c_k^{(h)}$  will be given later.
We identify $\left[\omega_L^{(h)}\right]$ as the
Poincar\'e dual to $D_L^{(h)}$,
i.e. $\left[\omega_L^{(h)}\right]=c_1\left([D_L^{(h)}]\right)$
where, as usual, $[D]$ denotes the line bundle associated to a given
divisor $D$ (see for example \cite{GH}) and $c_1$ denotes the first
Chern class.

We now fix the $c_k^{(h)}$'s by requiring that
$Z_h$'s defined in (\ref{2}) satisfy the recursion relations
(\ref{1}). Two facts are crucial to obtain recursion relations:
first,
in evaluating the relevant integrals
will appear
only the components of the boundary $\partial {\overline{\cal M}}_{h,2}$
of the form $\overline{\cal M}_{h-k,i}\times \overline{\cal M}_{k,j}$
with $i=j=2$ and $\overline{\cal M}_{h-1,4}$,
second, $\omega_{WP}^{(h,2)}$ satisfies the
restriction phenomenon mentioned above. In particular, considering the
natural embedding
\beq
i: {\overline{\cal M}_{k,2}}\to
{\overline{\cal M}_{k,2}}\times * \to
{\overline{\cal M}_{k,2}}\times {\overline{\cal M}_{h-k,2}}
\to \partial {\overline{\cal M}_{h,2}}\to {\overline{\cal M}_{h,2}},
\qquad h>k,
\label{shutupandplayyerguitar}\eeq
one has by \cite{Wolpert1}\cite{Wolpert}
\beq
\left[\omega_{WP}^{(k,2)}\right]=i^*
\left[\omega_{WP}^{(h,2)}\right],
\label{hotrats}\eeq
and similarly for $\left[\omega_{WP}^{(h-1,4)}\right]$.
By Poincar\'e duality one obtains
\begin{equation}
\int_{\overline{\cal M}_{h,2}}
\left(\omega_{WP}^{(h,2)}\right)^{3h-2}\wedge \omega_L^{(h)}=
\left[\omega_{WP}^{(h,2)}\right]^{3h-2}\cap \left[D^{(h)}_L\right],
\label{3}\end{equation}
so that by (\ref{shutupandplayyerguitar})-(\ref{hotrats}) it follows that
$$
Z_{h+1}=
{1\over (3h+2)!}\left\{c_0^{(h+1)}\left[\omega_{WP}^{(h,4)}\right]^{3h+1}
\cap \left[\overline{\cal M}_{h,4}\right]+\right.
$$
\beq
\left.\sum_{k=1}^{h}c_k^{(h+1)} \left[\omega_{WP}^{(h-k+1,2)}+
\omega_{WP}^{(k,2)}\right]^{3h+1}
\cap \left[\overline{\cal M}_{h-k+1,2}\times \overline{\cal
M}_{k,2}\right]\right\}.
\label{6}\end{equation}
The only contribution in the second term in the RHS comes from the
$\left(\omega_{WP}^{(h-k+1,2)}\right)^{3(h-k)+2}\wedge
\left(\omega_{WP}^{(k,2)}\right)^{3k-1}$ term, therefore we have
$$
Z_{h+1}=
{1\over (3h+2)!}
\left\{
c_0^{(h+1)}\int_{\overline{\cal M}_{h,4}}
\left(\omega_{WP}^{(h,4)}\right)^{3h+1}+\right.
$$
\beq
\left.\sum_{k=1}^{h}c_k^{(h+1)}
\left(^{3h+1}_{3k-1}\right)
\int_{\overline{\cal M}_{h-k+1,2}}
\left(\omega_{WP}^{(h-k+1,2)}\right)^{3(h-k)+2}
\int_{\overline{\cal M}_{k,2}}
\left(\omega_{WP}^{(k,2)}\right)^{3k-1}\right\}.
\label{7}\end{equation}

The recursion relations (\ref{7}) coincide
with (\ref{1}) if we set
\beq
c_0^{(h+1)}={(3h+2)\over 6}{W_h\over a_{h,4}}=
{(3h+2)\over 6}{25h^2-1\over 4 a_{n,4}}Z_h,\quad
c_k^{(h+1)} =-{3h+2\over 2}\left({Z_kZ_{h-k+1}\over a_{k,2} a_{h-k+1,2}}
\right),
 \qquad k>0,
\label{8}\eeq
where
$$
a_{k,n}={1\over (3k-3+n)!}\int_{\overline{\cal M}_{k,n}}
\left(\omega_{WP}^{(k,n)}\right)^{3k-3+n},
$$
is the Weil-Petersson volume of $\overline{\cal M}_{k,n}$
times $\left(1/\pi^2\right)^{3k-3+n}$ and $W_h$ has been defined in (\ref{m2}).
Notice that all the coefficients $c^{(h)}_k$ are rational numbers so
that $D_L$ defines a rational homology class and the above computations
can be interpreted in the sense of rational intersection theory.

Comparing with (\ref{0}) with (\ref{2}) we have asymptotically
\beq
{\cal Z}(t)\sim
e^{2i\Theta}t^{1\over 2}
+\sum_{h=1}^\infty {e^{2i\Theta(1-h)}t^{-{5\over 2}h+{1\over 2}}
\over (3h-1)!}
\int_{\overline{\cal M}_{h,2}}
\left(\omega_{WP}^{(h,2)}\right)^{3h-2}\wedge \omega_L^{(h)}.
\label{iuhdw}\eeq

If one is able to express the coefficients $Z_h$ in terms of the
expansion around
$t=0$  \cite{matone2}\cite{BMM1} then,
since the divisor of the expansions around $t=0$ and $t\to +\infty$
are known, it should be also
possible to understand in more detail the
reduction mechanism from genus $h$ to punctured spheres.
We remark that the structure used in the derivation of the recursion
relations can be generalized to integrals of $\overline{\cal M}_{h,n}$
yielding recursion relations for $n$-puncture correlator. Furthermore,
 together with a suitable
choice of the divisors at infinity, the use of the restriction phenomenon
seems to be
useful to
investigate intersection theory on $\overline{\cal M}_{h,n}$.

\vspace{0.5cm}

{\bf 4.} As in \cite{BMM1} one can construct an infinite
dimensional
moduli space of Riemann surfaces including those with infinitely many handles
via the following inductive limit. We consider the embedding
$$
i_h: {\overline{\cal M}_{h,2}}\to
{\overline{\cal M}_{h+1,2}}, \quad h>0,
$$
and for $q\in {\bf R}_+$ we define
\beq
\overline{{\cal M}}_{\infty,2}(q)=
\coprod_{h=1}^{\infty}
\left(
\overline{{\cal M}}_{h,2}\times [0,q^h]\right)/
\left(
\overline{{\cal M}}_{h,2},q^h\right)
\sim
\left(i_h\left(
\overline{{\cal M}}_{h,2}\right),0\right).
\label{gdjheuiekd}\end{equation}
Let $dy$ denote the Lebesgue measure on ${\bf R}$
 and define the indefinite rank form
\begin{equation}
{\cal T}_{\infty}=\sum_{h=1}^{\infty}
{e^{2i\Theta(1-h)}{\omega_{WP}^{(h,2)}}^{3h-2}\wedge\omega_L^{(h)}\over
(3h-1)!}\wedge dy,
\label{aaaaajkhncn}\end{equation}
then asymptotically
\begin{equation}
{\cal Z}(t)\sim
e^{2i\Theta}t^{1/2}
+t^{1/2}\int_{\overline{{\cal M}}_{\infty,2}(t^{-5/2})}
{\cal T}_{\infty}.
\label{wakayawaka}\end{equation}
One can give a meaning to the RHS of (\ref{wakayawaka})
in terms of a perturbation series. If the asymptotic series is Borel
summable, as presumably happens in 2D gravity with $\Theta$-vacua,
a more precise meaning can be given via Borel summation.

Equation (\ref{wakayawaka}), expresses
the asymptotic behaviour of the specific heat of
nonperturbative 2D quantum gravity as
an integral on an infinite dimensional space involving moduli spaces of
all double-punctured Riemann surfaces, and hence can be interpreted
in a sense as a kind of realization of the
Friedan-Shenker program \cite{FriedanShenker} in the asymptotic region.

\vspace{0.5cm}

{\bf 5.} We now give a conjectural argument that could relate our
{\it Ansatz} (\ref{2}) to the path-integral approach to Liouville
gravity. In this line of thought a key step is the
Duistermaat-Heckman (DH) \cite{DuistermaatHeckman} which roughly speaking
corresponds to the following statement. Let $X$ be a $2n$-dimensional
symplectic manifold with symplectic form $\omega$  and $H$ a Hamiltonian
on $X$. Then integrals such as
$$
{1\over n!}\int_X \omega^n e^{-H},
$$
only depend on the behaviour of the integrand near the critical points
of the flow of the Hamiltonian vector field.  The point is that in a
path-integral approach one expects that the contribution at genus
$h$ to the two-puncture correlation function of 2D gravity is given by
\beq
\langle \left({\cal O}_0\right)^2\rangle_h=
{1\over (3h-1)!}
\int_{\overline {\cal M}_{h,2}}\left(\omega_{WP}^{(h,2)}\right)^{3h-1}e^{-H},
\label{qkudgw}\eeq
where $H$ is an ``effective action''
arising from the integration in the path-integral
at fixed moduli in
$\overline {\cal M}_{h,2}$. The two-form $\omega_{WP}^{(h,2)}$
is symplectic on $\overline {\cal M}_{h,2}$, regular in the interior
and extending as a current to the boundary, therefore,
 regarded as a map from
$T^* \overline {\cal M}_{h,2}$ to
$T \overline {\cal M}_{h,2}$,
$\left(\omega_{WP}^{(h,2)}\right)^{-1}$ has zeroes only on
$\partial\overline {\cal M}_{h,2}$. Furthermore, since
$\omega_{WP}^{(h,2)}$ is a K\"ahler form, the Hamiltonian vector field
is given by $\left(\omega_{WP}^{(h,2)}\right)^{-1}dH$ so that
the flow of the Hamiltonian vector field has critical points at
$\partial\overline {\cal M}_{h,2}$. Let us assume that DH applies
to the integral
(\ref{qkudgw}) and furthermore it gets contribution only
from the critical points in the component of
$\partial\overline {\cal M}_{h,2}$ whose
factor contain an even number of punctures.
Then one expects
$$
\langle \left({\cal O}_0\right)^2\rangle_{h+1}\sim
\alpha_{h+1}{1\over (3h+1)!}
\int_{\overline {\cal M}_{h,4}}
 \left(\omega_{WP}^{(h,4)}\right)^{3h+1}
e^{-H}+
$$
$$
\beta_{h+1}\sum_{k=1}^h\left[{1\over [3(h-k)+2]!}
\int_{\overline {\cal M}_{h-k+1,2}}
 \left(\omega_{WP}^{(h-k+1,2)}\right)^{3(h-k)+2}
e^{-H}\right]\cdot
$$
\beq
\left[
{1\over (3k-1)!}
\int_{\overline {\cal M}_{k,2}}
 \left(\omega_{WP}^{(k,2)}\right)^{3k-1}
e^{-H}\right]=
\alpha_{h+1}\langle \left({\cal O}_0\right)^4\rangle_h+
\beta_{h+1}\sum_{k=1}^h
\langle \left({\cal O}_0\right)^2\rangle_{h-k+1}
\langle \left({\cal O}_0\right)^2\rangle_{k},
\label{joegarage}\eeq
where $\alpha_{h+1}$, $\beta_{h+1}$ are possibly $t$-dependent coefficients.
One can remark the analogy with the recursion relations derived from KdV
in topological gravity.
Let us introduce
the cohomology classes
$\left[\eta^{(h,k)}\right]\in H^{2}\left({\overline{\cal
M}_{h,2}}\right)$, $k=0,\ldots, h$, Poincar\'e dual
of
$\overline {\cal M}_{h-1,4}$,
 $\overline {\cal M}_{h-k,2}\times \overline {\cal M}_{k,2}$,
$k=1,\ldots, h$.
Introducing the normalized Weil-Petersson volumes $a_{k,n}$
and using (\ref{m2}),
one obtains for the asymptotic behaviour of the correlations
$$
\langle \left({\cal O}_0\right)^2\rangle_{h+1}\sim
{1\over (3h+2)!}
\int_{\overline {\cal M}_{h+1,2}}
 \left(\omega_{WP}^{(h+1,2)}\right)^{3h+1}\wedge\left[
\alpha_{h+1} {25h^2-1\over 4}
{\langle \left({\cal O}_0\right)^2\rangle_{h}\over a_{h,4}}
\eta^{(h+1,0)}+
\right.
$$
\beq
\left.\beta_{h+1}\sum_{k=1}^h
{\langle \left({\cal O}_0\right)^2\rangle_{h-k+1}\over
a_{h-k+1,2}} {\langle\left({\cal O}_0\right)^2\rangle_{k}\over
a_{k,2}}\eta^{(h+1,k)}\right].
\label{joegarage2}\eeq
Since the asymptotic expression of
$\langle\left({\cal O}_0\right)^2\rangle_{h}$ evaluated at $t=1$ is
equal to $-Z_h$, by setting
$t=1$ in
(\ref{joegarage2}) we derive
\beq
Z_{h}\sim
{1\over (3h-1)!}
\int_{\overline {\cal M}_{h,2}}
 \left(\omega_{WP}^{(h,2)}\right)^{3h-2}\wedge\omega^{(h)},
\label{gdteb}\eeq
with $\omega^{(h)}$ a two-form given by
\beq
\omega^{(h+1)}=\alpha_{h+1}{25h^2-1\over 4} {Z_h\over a_{h,4}}
\eta^{(h+1,0)}-\beta_{h+1}\sum_{k=1}^h
{Z_{h-k+1}\over a_{h-k+1,2}}
{Z_{k}\over a_{k,2}}\eta^{(h+1,k)},
\label{enormous3}\eeq
exhibiting the same structure of the two-form $\omega_L^{(h+1)}$
introduced in the {\it Ansatz} (\ref{2}) with the coefficients
$c_k^{(h+1)}$ given by (\ref{8}).

\vspace{0.5cm}

{\bf 6.}
The above results can be seen as a first step to recover the full structure
of Liouville theory. To see this note that divisors at infinity of
moduli spaces are related to
determinant line
bundles $\lambda_k$ and Weil-Petersson two-form $\omega_{WP}$. For example in
$\overline {\cal M}_h$ one has for $h>2$ \cite{Wolpert1}
\beq
c_1(\lambda_H)={1\over 24}[\omega_{WP}/\pi^2]+{1\over 12}D_0+{1\over 24}D_1+
{1\over 12}D_2+
\cdot\cdot\cdot +{1\over 12}D_{[h/2]},
\label{mm1}\eeq
where $\lambda_H\equiv \lambda_1$ is the Hodge bundle.
 The structure (\ref{mm1}) together with the Mumford
isomorphism $\lambda_k\cong \lambda_1^{6k^2-6k+1}$
are basic in the description of the
Polyakov volume form $d\pi_h$ of the critical bosonic string
\cite{BelavinKnizhnik}\cite{BeilinsonManin}
\beq
Z=\int_{\overline {\cal M}_h}d\pi_h,
\label{ihx}\eeq
where $d\pi_h=e^{\beta(2-2h)}|\mu|^2$, with $\mu$ the Mumford form.
The Weil-Petersson volume form is associated
to the ghost zero modes whereas sections of determinant line bundles
$\lambda_k$ arise from path integral on $b$-$c$ systems of weight
$k$. In the description of the algebraic-geometrical structure
of the specific heat we recovered the divisors at infinity
and the Weil-Petersson two-form. Generalizing the structure
(\ref{mm1}) to the case of $\overline{\cal M}_{h,n}$ one could expect
to be  able to describe the cohomology structure of $Z_h$ directly in terms
of Weil-Petersson two-form and determinant line bundles, and then
to recover the structure of the Liouville path-integral, problem
pioneered in \cite{1},
associated to matrix models.
In other words this suggests that it could be possible to reconstruct
the full structure of the theory starting from nonperturbative
results and then going back
to the continuum formulation.

\end{document}